\documentclass[%
 reprint,
superscriptaddress,
 amsmath,amssymb,
 aps,
]{revtex4-2}
\usepackage{hyperref}
\usepackage{graphicx}
\usepackage{dcolumn}
\graphicspath{{Figures/PNG/}{Figures/PDF/}{Figures/}}

\usepackage{bm}

\begin{document}

\preprint{APS/123-QED}

\title{All-silicon quantum light source by embedding an atomic emissive center in a nanophotonic cavity}

\author{Walid Redjem}
 \altaffiliation{Equal contribution}

\affiliation{Department of Electrical Engineering and Computer Sciences, University of California Berkeley, Berkeley, CA 94720, USA}

\author{Yertay Zhiyenbayev}
 \altaffiliation{Equal contribution}
\affiliation{Department of Electrical Engineering and Computer Sciences, University of California Berkeley, Berkeley, CA 94720, USA}

\author{Wayesh Qarony}
 \altaffiliation{Equal contribution}
\affiliation{Department of Electrical Engineering and Computer Sciences, University of California Berkeley, Berkeley, CA 94720, USA}

\author{Vsevolod Ivanov}
\affiliation{Accelerator Technology and Applied Physics Division, Lawrence Berkeley National Laboratory, Berkeley, California 94720, USA}

\author{Christos Papapanos}
\affiliation{Department of Electrical Engineering and Computer Sciences, University of California Berkeley, Berkeley, CA 94720, USA}

\author{Wei Liu}
\affiliation{Accelerator Technology and Applied Physics Division, Lawrence Berkeley National Laboratory, Berkeley, California 94720, USA}

\author{Kaushalya Jhuria}
\affiliation{Accelerator Technology and Applied Physics Division, Lawrence Berkeley National Laboratory, Berkeley, California 94720, USA}

\author{Zakaria Al Balushi}
\affiliation{Department of Materials Science and Engineering, University of California Berkeley, Berkeley, California 94720, USA}
\affiliation{Materials Sciences Division, Lawrence Berkeley National Laboratory, Berkeley, California 94720, USA}

\author{Scott Dhuey}
\affiliation{Molecular Foundry, Lawrence Berkeley National Laboratory, Berkeley, California 94720, USA}

\author{Adam Schwartzberg}
\affiliation{Molecular Foundry, Lawrence Berkeley National Laboratory, Berkeley, California 94720, USA}

\author{Liang Tan}
\affiliation{Molecular Foundry, Lawrence Berkeley National Laboratory, Berkeley, California 94720, USA}

\author{Thomas Schenkel}
\affiliation{Accelerator Technology and Applied Physics Division, Lawrence Berkeley National Laboratory, Berkeley, California 94720, USA}

\author{Boubacar Kanté}
 \email{Corresponding author: bkante@berkeley.edu}
\affiliation{Department of Electrical Engineering and Computer Sciences, University of California Berkeley, Berkeley, CA 94720, USA}
\affiliation{Materials Sciences Division, Lawrence Berkeley National Laboratory, Berkeley, California 94720, USA}
\date{\today}

\begin{abstract}
Silicon is the most scalable optoelectronic material, and it has revolutionized our lives in many ways. The prospect of quantum optics in silicon is an exciting avenue because it has the potential to address the scaling and integration challenges, the most pressing questions facing quantum science and technology. We report the first all-silicon quantum light source based on a single atomic emissive center embedded in a silicon-based nanophotonic cavity. We observe a more than 30-fold enhancement of luminescence, a near unity atom-cavity coupling efficiency, and an 8-fold acceleration of the emission from the quantum center. Our work opens avenues for large-scale integrated all-silicon cavity quantum electrodynamics and quantum photon interfaces with applications in quantum communication, sensing, imaging, and computing.
\end{abstract}

\maketitle

Quantum science and technologies promise to revolutionize our societies \cite{nielsen_quantum_2002,macfarlane_quantum_2003}. In the search for the ideal quantum information processing platform, “scaling” is perhaps the most challenging question due to the fundamental but contradictory requirements for quantum systems to simultaneously be isolated and controllable from the environment in large arrays of interacting qubits \cite{de_leon_materials_2021,arute_quantum_2019}. Among many quantum information platforms ranging from superconducting qubits to trapped ions, quantum photons play a fundamental role because they are necessary for future quantum networks to enable communication between distant quantum nodes \cite{bouwmeester_experimental_1997,clarke_superconducting_2008,devoret_superconducting_2013,monroe_scaling_2013,kimble_quantum_2008}. Quantum photons have been generated from an extensive range of platforms, including quantum dots, color-centers in diamonds such as NV, SiV, and SnV, or defects in two-dimensional materials such as hBN \cite{santori_indistinguishable_2002,gruber_scanning_1997,lee_deterministic_2014,hayee_revealing_2020,xu_creating_2021}. The scaling challenge is currently being addressed using hybrid material platforms and metamaterials in which quantum light sources are optimized and integrated into more complex scalable systems, following the example of heterogeneous integration in the classical domain \cite{elshaari_hybrid_2020,wan_large-scale_2020,carter_quantum_2013,santiago-cruz_resonant_2022}. However, the challenge for integrating quantum devices is more significant than for classical systems because each interface allows losses and decoherence that need to be minimized. It is thus fundamental to minimize the number of interfaces by deeply integrating intrinsically scalable platforms.

Silicon is currently the most scalable optoelectronic material. Despite the lack of efficient classical light sources based on silicon, emissive centers have been observed in silicon since the end of the 1980s \cite{davies_optical_1989}. It is only during the last two years that single centers in silicon have been isolated \cite{redjem_single_2020,hollenbach_engineering_2020, durand_broad_2021,higginbottom_optical_2022}. Since then, emissive centers in silicon have been coupled to waveguides, and more recently, an ensemble of centers has been integrated into ring resonators \cite{prabhu_individually_2022,deabreu_waveguide-integrated_2022,komza_indistinguishable_2022,lefaucher_cavity-enhanced_2022}. However, deterministic single photon sources based on silicon emissive centers have remained elusive due to the lack of controlled manufacturing approaches and the complexity of materials interfaces after device fabrication. We report the first all-silicon quantum light source based on an atomic emissive center in a silicon nanophotonic cavity. The manufacturing of the centers in silicon-on-insulator substrates, with controlled densities and dipole orientations, enables their overlap with designed nanophotonic cavities. We demonstrate the successful alignment of a quantum defect and nanophotonic cavity dipole moments and tune the nanophotonic cavity to overlap its resonance with the zero-phonon line of the silicon-based quantum defect. We achieve a more than 30-fold enhancement of the luminescence intensity and an 8-fold acceleration of the single photon emission rate. Our results open the door to large-scale integrated all-silicon quantum optics devices and systems for applications in quantum communication, sensing, imaging, and computing.

\begin{figure*}[p]
    \centering
    \includegraphics[width=0.8\textwidth]{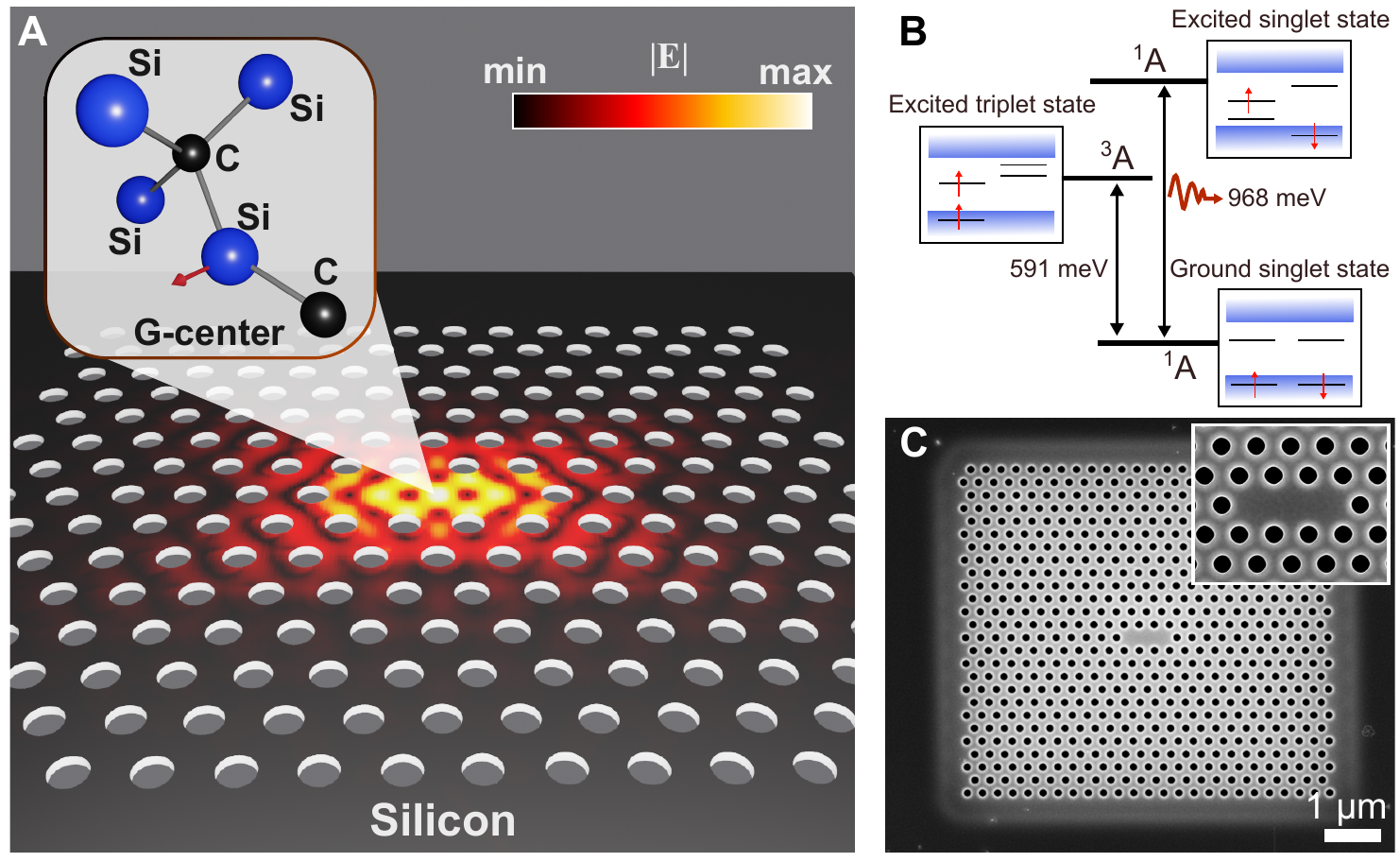}
    \caption{\textbf{A single atomic emissive center embedded in a silicon-photonic cavity.}  (\textbf{A}) Silicon quantum interface with an “atomic defect” located within the “photonic defect” cavity of three missing holes in a triangular photonic crystal (PhC). The atomic defect is the G-center in silicon made of two substitutional carbon atoms (black spheres) bound to the same silicon self-interstitial (blue sphere). The red arrow indicates the direction of the dipole moment of the G-center. The G-center is one of a broad diversity of recently observed emissive centers in silicon, the most scalable optoelectronic material. The computed electromagnetic mode of the cavity is superimposed on the sketch of the PhC, evidencing the high confinement of the electromagnetic field in the region of missing holes in the triangular lattice. The cavity is fabricated so that its dipole moment and the dipole moment of the defect are colinear (see Supplementary Materials). The electric field strength peaks at the center of the cavity and exponentially decays in the bulk of the PhC. (\textbf{B}) Energy level diagram of the G-center in silicon comprising a ground singlet state, a dark excited triplet state, and an excited singlet state. The cavity can be tuned to be in resonance with the radiative transition between the excited and ground singlet states to enhance light-matter interaction. (\textbf{C}) Scanning electron microscope (SEM) image of a fabricated silicon-based atom-cavity system suspended in the air. The successful embedding of a single G-center in a photonic cavity involved a controlled sequence of fabrication steps using a commercial 230 nm thick silicon-on-insulator (SOI) wafer that is carbon implanted, followed by electron beam lithography, dry etching, thermal annealing, and wet etching (see Supplementary Materials). The fabrication steps, compatible with standard CMOS processes, are optimized to increase the probability of single color centers in cavities (see Supplementary Materials).}
    \label{fig:fig1}
\end{figure*}

The proposed all-silicon atom-cavity system, presented in Fig. \ref{fig:fig1}A, consists of a single defect in silicon embedded in a photonic crystal (PhC) defect cavity. The PhC cavity consists of three missing holes in a suspended triangular lattice of holes \cite{sakoda_optical_2005}. The atomic defect, the G-center in silicon, is made of two substitutional carbon atoms (black spheres) bound to the same silicon self-interstitial (blue sphere). The manufacturing process starts with the implantation of carbon ($^{13}$C) with an energy of 36 keV in a commercial 230 nm thick silicon-on-insulator (SOI) wafer. The implantation is followed by electron beam lithography, dry etching, thermal annealing, and wet etching (see Supplementary Materials). Secondary ion mass spectroscopy (SIMS) measurements indicate that the implanted carbon and the atomic centers created during the annealing process are located in the middle of the silicon layer (see Supplementary Materials). 

The dipole moment of the center is computed by density functional theory \cite{ivanov_effect_2022}. It is in the plane as indicated by the red arrow in the inset of Fig.\ref{fig:fig1}A.  The G-center is one of a broad diversity of recently observed emissive centers in silicon and its electronic structure, presented in Fig. \ref{fig:fig1}B, comprises a ground singlet state, a dark excited triplet state, and an excited singlet state \cite{udvarhelyi_identification_2021}. The computed electromagnetic mode of the cavity for the transverse electric polarization is superimposed on the sketch of the PhC, evidencing the high confinement of the electromagnetic field in the region of missing holes in the triangular lattice. This polarization matches the orientation of the atomic defect dipole moment. The deterministic positioning of atomic-scale defects in photonic cavities has been challenging for most platforms and has not yet been achieved for silicon-emissive centers. It requires not only overlap of the quantum defect with highly confined optical modes but also the alignment of the dipole moments of the atom and the cavity. To overcome this challenge in our platform, we first investigated the scalable manufacturing of single emissive centers with controllable densities and inhomogeneous broadening. We identified an annealing time window below which only ensembles of centers are created and beyond which all single centers are destroyed (see Supplementary Materials). We also find that shorter annealing time within that window minimizes the inhomogeneous broadening of the zero-phonon line (ZPL) of the quantum emitters, a critical requirement for overlapping the ZPL with a designed nanophotonic resonance to enhance light-matter interaction (see Supplementary Materials). The controlled density and inhomogeneous broadening of quantum centers increase the probability of overlap with an array of finite-size photonic crystal cavities. We subsequently investigated the polarization response of created emissive centers, and a statistical analysis presented in the supplementary materials indicates a preferential orientation of the emitters in silicon. We then fabricate PhC cavities so that the dipole moments of the cavities and centers align. Fig. \ref{fig:fig1}C presents a scanning electron microscope (SEM) image of a fabricated silicon-based atom-cavity system. The inset presents the cavity with a mode volume of $0.66(\lambda_{\text{cav}}/n)^3$. The successful embedding of a single center in a cavity involved a controlled sequence of CMOS-compatible fabrication steps.

\begin{figure*}[p]
    \centering
    \includegraphics[width=0.8\textwidth]{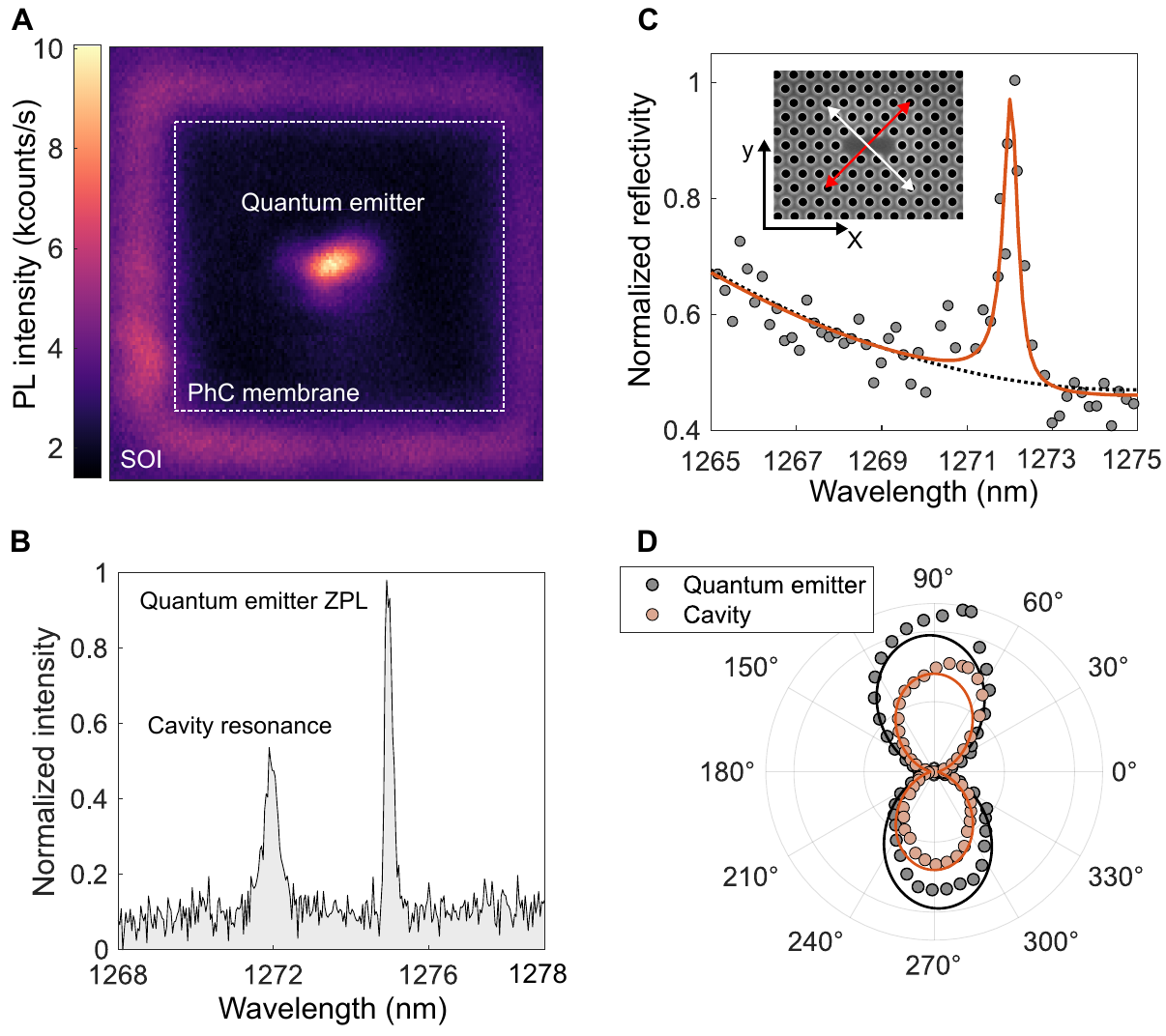}
    \caption{\textbf{Experimental characterization of the silicon-based quantum emitter and cavity.} (\textbf{A}) Photoluminescence (PL) raster scan of a device with a single emitter in the cavity. The boundary of the finite photonic crystal (PhC) is indicated by the dashed white line and the suspended PhC is surrounded by the silicon-on-insulator (SOI) wafer. The photoluminescence signal shows bright emission from a color center within the boundaries of the cavity. (\textbf{B}) The photoluminescence of the photonic device exhibits a sharp peak at ~1275 nm and a blue-shifted broader peak at ~1272 nm, corresponding to the zero-phonon line of the color-center and the cavity resonance, respectively. (\textbf{C}) Reflectivity of the photonic crystal cavity obtained by resonant scattering measurements. The cavity is illuminated with a linearly polarized white light source (white arrow) at 45 degrees with respect to the cavity axis that is along the X-direction.  The polarized signal perpendicular to the excitation is collected (red arrow) to probe the cavity mode. (\textbf{D}) The polarization diagram of the cavity mode detuned from the ZPL is shown in orange. The polarization diagram of a quantum emitter alone is shown in black. The polarizations agree well with a dipolar model (solid lines) and are well aligned (see Supplementary Materials).}
    \label{fig:fig2}
\end{figure*}

Fig. \ref{fig:fig2}A presents the photoluminescence (PL) raster scan of a device with bright emission from a color center within the boundaries of the cavity. The dashed white line indicates the boundary of the finite PhC, and the suspended PhC is surrounded by the silicon-on-insulator (SOI) wafer. The photoluminescence of the photonic device, presented in Fig. \ref{fig:fig2}B, exhibits a sharp peak at ~1275 nm and a blue-shifted broader peak at ~1272 nm, corresponding to the ZPL of the color-center and the cavity resonance, respectively. The cavity is further characterized in reflectivity using resonant scattering measurements in Fig. \ref{fig:fig2}C. The cavity is illuminated with a linearly polarized white light source (white arrow) at 45 degrees with respect to the cavity axis that is along the X-direction.  The signal polarized perpendicular to the excitation is collected (red arrow) to probe the cavity mode, and resonance is observed at ~1272 nm, in perfect agreement with the PL measurement. The reflectivity is fitted with a Fano resonance line shape giving an intrinsic quality factor ($Q$) of 3209. The experimental value is comparable to the theoretical $Q$ of 6000 and the discrepancy is attributed to fabrication imperfections. Fig. \ref{fig:fig2}D presents the polarization diagram of the cavity mode detuned from the ZPL in orange and the polarization diagram of a quantum emitter alone in black. The polarizations agree well with a dipolar model (solid lines) and have been successfully aligned.

\begin{figure*}[p]
    \centering
    \includegraphics[width=0.8\textwidth]{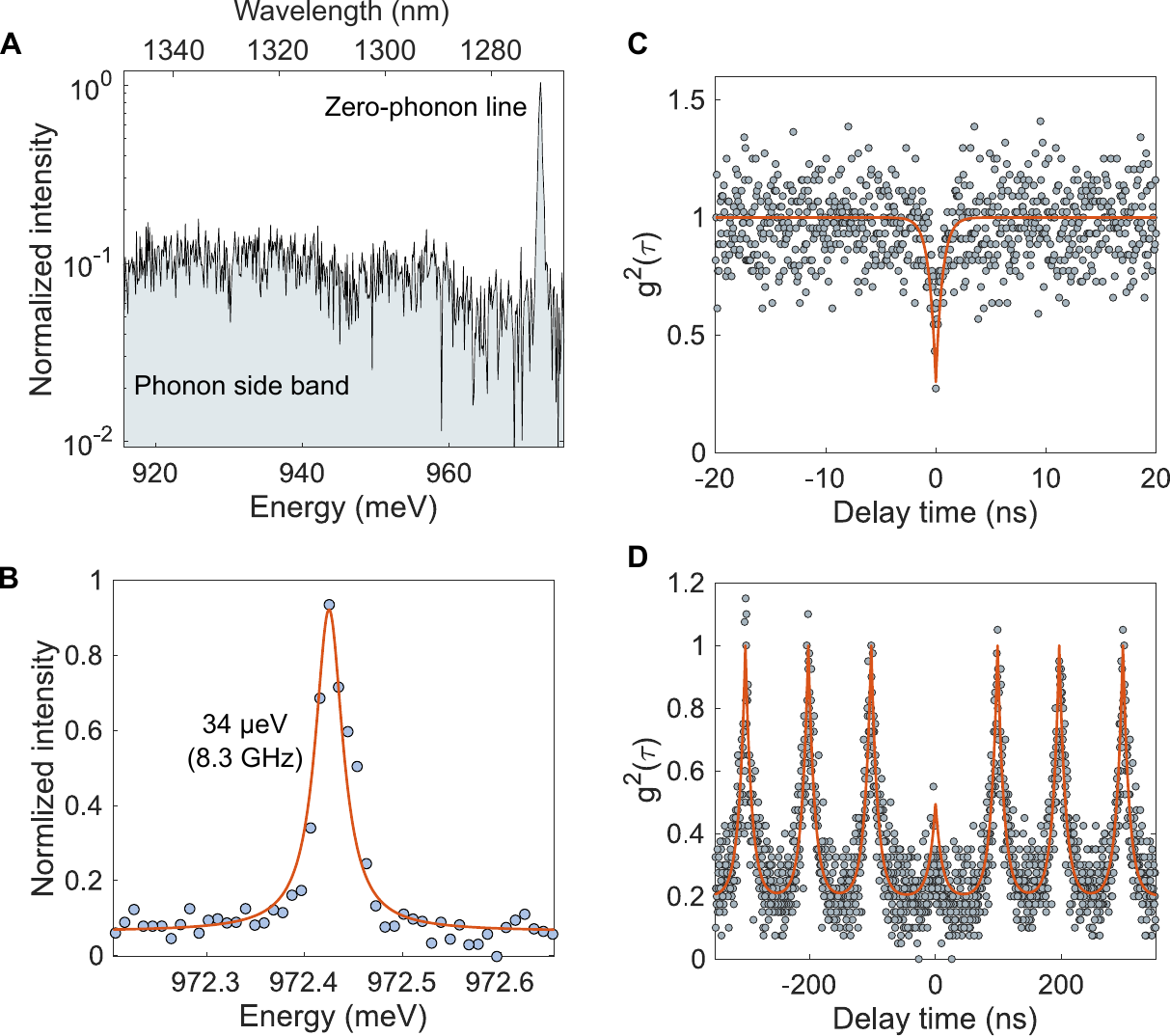}
    \caption{\textbf{Quantum coherence measurements of the emitter in the cavity.} (\textbf{A}) Spectrum of the quantum emitter over a broad range of energy showing the zero-phonon line (ZPL) of the silicon emissive center and its phonon sideband. (\textbf{B}) Spectrum of the quantum emitter using a high-resolution grating. The ZPL is located at 972.43 meV and has a linewidth of 8.3 GHz.  (\textbf{C}) Second-order autocorrelation measurements of the emission from the cavity under continuous excitation. The antibunching at zero delay confirms the successful spatial overlap of a single silicon emissive center with the nanophotonic cavity with an antibunching at zero delay $g^{(2)}(0) = 0.30 \pm 0.07$. (\textbf{D}) Second-order autocorrelation measurements under pulsed excitation at a repetition rate of 10 MHz demonstrating on-demand single-photon generation from the all-silicon platform. Autocorrelation measurements are performed using a Hanbury-Brown and Twiss interferometer with superconducting nanowire single-photon detectors (see Supplementary Materials).}
    \label{fig:fig3}
\end{figure*}

In Fig. \ref{fig:fig3}A, the spectrum of the quantum emitter over a broad range of energy shows the zero-phonon line (ZPL) of the silicon emissive center and its phonon sideband. Fig. \ref{fig:fig3}B presents the spectrum of the quantum emitter using a high-resolution grating. The ZPL is located at 972.43 meV and has a linewidth of 8.3 GHz. To demonstrate that the bright emission from the middle of the cavity corresponds to a single emissive center, we performed quantum coherence measurements of the emitter in the cavity. Autocorrelation measurements, shown in Fig. \ref{fig:fig3}C, are performed using a Hanbury-Brown and Twiss interferometer with superconducting nanowire single-photon detectors (see Supplementary Materials). The second-order correlation measurements of the emission from the cavity under continuous excitation exhibit an antibunching, confirming the successful spatial overlap of a single silicon emissive center with the nanophotonic cavity with an antibunching at a zero delay $g^{(2)}(0) = 0.30 \pm 0.07$. Autocorrelation measurements under pulsed excitation at a repetition rate of 10 MHz are presented in Fig. \ref{fig:fig3}D and they demonstrate on-demand single-photon generation from the all-silicon platform.

\begin{figure*}[p]
    \centering
    \includegraphics[width=0.8\textwidth]{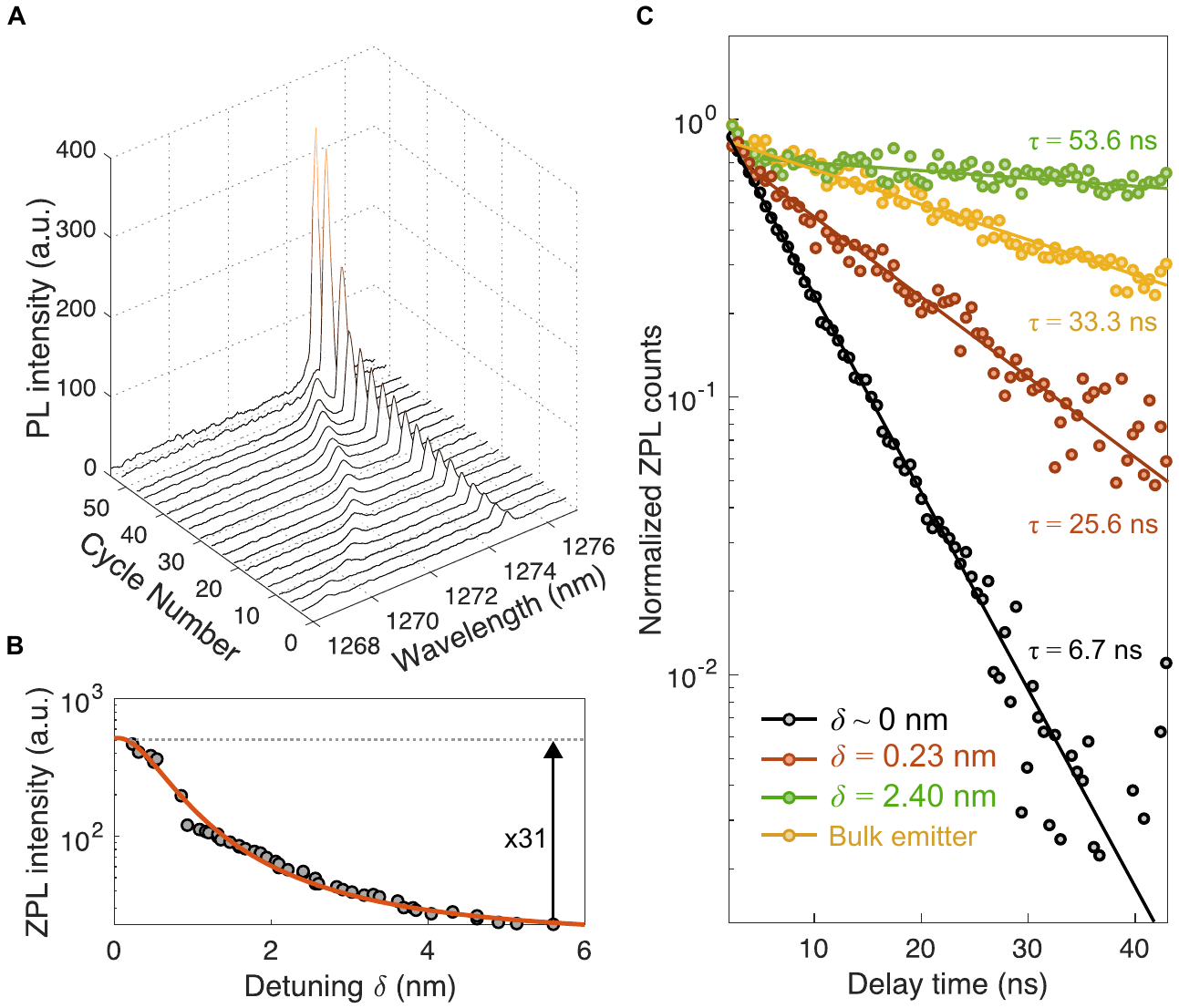}
    \caption{\textbf{Spectral tuning of the nanocavity and enhanced atom-cavity interaction.} (\textbf{A}) Cavity tuning as a function of the argon gas injection cycle. The enhancement of the single center in the cavity requires spatial and spectral overlap. Spatial overlap was achieved in Fig. \ref{fig:fig2} and Fig. \ref{fig:fig3}  and spectral overlap is achieved here by tuning the nanophotonic cavity. Gas injection modifies the effective index of cavity mode and tunes the resonance wavelength of the cavity that is shifted from $\sim$1269 nm to $\sim$1275 nm. As the cavity resonance is shifted towards the ZPL of the quantum center, the photoluminescence is enhanced to reach a maximum at $\sim$1275 nm, where the spectral overlap is achieved. (\textbf{B}) Zero-phonon line intensity as a function of the cavity detuning. An enhancement larger than 30 is achieved on resonance. (\textbf{C}) Excited lifetime for cavity detuning of $\delta$ = 2.40 nm, $\delta$ = 0.23 nm, and $\delta$ = 0.00 nm. The lifetime of emitted photons shortens from 53.6 ns to 6.7 ns when the detuning between the cavity and emitter is decreased. An 8-fold reduction in the lifetime is experimentally observed when the overlap is achieved compared to the off-resonance case. These results constitute the first all-silicon quantum light source using a silicon emissive center in a cavity, and the center can be further accelerated by designing cavities with higher quality factors as well as more deterministic positioning methods to further improve the emitter-cavity spatial overlap. The results will enable all-silicon quantum optics interfaces with silicon-emissive centers for scalable quantum optics.}
    \label{fig:fig4}
\end{figure*}

The enhancement of the single center in the cavity requires spatial and spectral overlap. Spatial overlap was achieved in Fig. \ref{fig:fig2} and Fig. \ref{fig:fig3}. To achieve spectral overlap, the nanophotonic cavity is tuned using cycles of argon gas injection. The injected gas condensates at the surface of the PhC and modifies the effective index of cavity mode tuning the resonance wavelength of the cavity that is shifted from ~1269 nm to ~1275 nm. In Fig. \ref{fig:fig4}A, as the cavity resonance is shifted towards the ZPL of the quantum center, the photoluminescence is enhanced to reach a maximum at ~1275 nm, where the spectral overlap is achieved. In Fig. \ref{fig:fig4}B, the ZPL intensity as a function of the cavity detuning shows an enhancement larger than 30 achieved on resonance. For cavity detuning varying from $\delta$ = 2.40 nm to $\delta$ = 0.00 nm, the excited lifetime shortens from 53.6 ns to 6.7 ns. An 8-fold reduction in the lifetime is experimentally observed when the overlap is achieved compared to the off-resonance case. Light-matter interaction in cavities is usually quantified using the Purcell factor ($F_p$) that measures the decay rate enhancement of the atom from free space to the cavity ($\gamma_{\text{cav}} = F_p\gamma_0$). It can be estimated by $F_p = (\tau_{\text{bulk}}/\tau_{\text{on}}-\tau_{\text{bulk}}/\tau_{\text{off}})/\eta$ where $\tau_{\text{bulk}}$ is the lifetime of a quantum emitter outside the PhC (dark yellow dots in Fig. \ref{fig:fig4}C), and $\tau_{\text{off}}$ is the lifetime for a detuning of 2.4 nm. The lifetime measured off-resonance is slightly longer than the one in the bulk because of the reduced density of state in the PhC gap \cite{yablonovitch_inhibited_1987}.  The percentage of photons emitted at the ZPL wavelength of the emitter (Debye-Waller) is $\eta = 15\%$, which was measured by comparing the count rate with and without the ZPL bandpass filter. The experimental Purcell factor of the defect cavity is $F_p \sim 29.0$. The coupling efficiency of the center to the cavity mode ($\beta$ factor) can be estimated by $1/\tau_{on}/[1/\tau_{on}+1/\tau_{off}]$, yielding a value of $\beta \sim 89\%$.

We thus reported the first all-silicon quantum light source based on an atomic emissive center in a silicon-based nanophotonic cavity. The quantum light source in silicon is one of a broad diversity of recently discovered centers that we successfully embedded in a silicon-on-insulator photonic crystal cavity. The quantum center is manufactured in silicon using a sequence of CMOS compatible steps that control the density, inhomogeneous broadening, and orientation of the dipolar moment of the emitters, enabling their efficient overlap with designed nanophotonic cavities. The performance of the quantum light source can be further improved by developing cavities with higher quality factors as well as more deterministic positioning methods to further improve the emitter-cavity spatial overlap. Our results will enable all-silicon quantum optics interfaces with silicon-emissive centers for scalable and integrated quantum optoelectronics.

\vbox{}                                 
\typeout{}                              
\bibliography{references}



\end{document}